# Subject Access through Community Partnerships:  A Case Study

Patricia A. Kreitz and Travis C. Brooks

Stanford Linear Accelerator Center, Stanford University
Stanford, California 94309

**Abstract**
Innovations in scholarly communication have resulted in changing roles for authors,
publishers and libraries. Traditional roles are disappearing and players are actively
seeking or reluctantly assuming new roles.  Library roles are changing as they become
involved in building and indexing electronic (e-) repositories and support new modes of
e-research.  A library-run service, the SPIRES particle physics databases, has not only
weathered, but also lead, many of the transitions that have shaped the landscape of e-
publishing and e-research.  This has been possible through an intense and in-depth
partnership with its user community.  The strategies used and lessons learned can help
other libraries craft cost-effective roles in this new environment.

Work supported by Department of Energy contract DE-AC03-76SF00515

# Introduction

The rise of the Web combined with a growing ease of writing and publishing electronically have begun a revolution in scholarly publishing and communication. Profoundly transformative innovations such as arXiv.org plus internet indexing and retrieval software such as Google™ are changing the academic information landscape. One area with the greatest potential for change is in the traditional processes and the players involved in providing access to the scholarly literature. In an e-research world, how will scholars be able to have persistent, useful, accurate, and timely access to that subset of the scholarly literature which is relevant to them? How will the roles of the players in the scholarly communication process evolve in the emerging e-publishing and e-researching world? Will authors become publishers and catalogers as well? Will journal publishers completely replace libraries or abstracting and indexing services? Will libraries, in turn, extend their roles both 'backwards' into the publication process and 'forwards' into more comprehensive subject access?

Brian Hawkins, past-president of Educause, observed that "There is no clear and defined role for libraries with regard to the digital resources accessible through the Net".[1] He challenges libraries to find a way to provide free and open searchability to the 'deep Web' of scholarly disciplines using a combination of software and humanware and judicious collaboration and partnering. How can libraries take such a leadership role? In this paper we look at some of the ways this might be achieved, particularly by involving authors and libraries in new partnerships as traditional roles change. We examine an active model of this new partnership, looking at how one library-managed system has worked with users to cost-effectively provide useful, persistent, accurate, and relevant access to the subset of particle physics scholarly literature.

# Changing Roles

Researchers, now known as 'content producers,' whose past role was to write articles and books that they handed off to other players in the publishing structure, have many more options today. They are often bypassing traditional publishers and are now self-publishing or publishing to an institutional archive or a subject-based archive. Reflecting the beginning stages of this revolution, many authors are choosing to combine new publication methods with traditional ones. The publication process for authors will continue to evolve in new ways as traditional copyright control is redefined and as institutional or subject-based repositories become part of the mainstream.

This trend of scholars retaining more control over their documents is being extended into the arena of subject access through a variety of experiments.[2] Some projects currently underway are testing the expectation that researchers should self-catalog their own works at the point of self-publishing them.[3] Creating the access points and indexing terms for their own works would replace the need for what had been in the past the 'down-stream' cataloging or metadata creation functions that have traditionally been performed by libraries or abstract/indexing services.

Having authors index their own works is appealing both intellectually, because researchers know their own writings best, and practically, because this would avoid the inevitable delays caused by relying on third party indexing. But are authors willing and able to assume this new role? Academic authors typically want to focus on research and teaching and may be extremely reluctant to become 'lifetime catalogers' of their written materials.[4] Will authors create sufficient metadata so that nothing will be needed but a 'Google™ on steroids'? We are too early in this revolution to know what new paths will be taken, but it is certain that the wealth of experimentation that is taking place will alter radically the traditional roles authors have played in the past.

Publishers as well are finding and taking advantage of new opportunities created by the upheaval in research communication. Both commercial and scholarly publishers have traditionally provided reviewing, editing, copyright control over, and persistent access to, scholarly works. They are now analyzing what roles they may play in a world where the act of publishing is no longer a single event, frozen in time, but distributed and dynamic, and the power of exclusive copyright control is weakening.[5] They are also pondering the effect that some of the changes already well underway may have on their enterprise. In particular, electronic repositories, inspired by the first e-print archive, arXiv.org, have now grown into adolescence. Not only have other subject-based e-print servers been started, but universities such as CalTech and MIT and collaborations of scholars such as the Public Library of Science are building an electronic-based role in publishing by creating successful digital repositories of scholarly works. These structured electronic repositories are now enabling the dissemination of a scholarly work, a traditional role of a publisher, to be separated from all other parts of the scholarly communication process, including 'publication'.[6]

Publishers may focus solely on their refereeing and certification role.[7] But if they do, how can they generate sufficient revenue? Will institutions, universities, or authors be willing to pay publishers sufficient income for the cachet of inclusion of a particular article in a journal? Not willing to trust their existence and revenue stream to a somewhat beleaguered publication function, journal publishers are experimenting with value-added services that extend their roles in new directions. Many are assuming that they will continue to provide electronically based peer reviewed scholarly publications, but also are contemplating a variety of other content, tools, and access systems that could be pay-per-use.[8] Will they try to build the kinds of products and services that have, in the past, been the responsibility of other parts of the scholarly communication process? Recent developments such as IOP's *BEC Matters!* portal [9] and Elsevier's SocSciNet.com[10] are expanding traditional journal publisher activities into subject searchable databases built, at present, on their own suite of products but attempting to reach beyond them to broader subject access through subfield Web portals. Will such experiments grow to a point where journal publishers control the access to and mining of their electronic full text resources so fully that they completely replace libraries or abstracting/indexing services?[11] If libraries allow this trend to continue, might subject access through publishers soon present the same pitfalls and monopolies that have plagued libraries in the current journal system?

**Library Challenges**

While new roles for authors and publishers are clearly of interest to libraries, foremost in the minds of libraries and librarians are the new roles that they themselves will be asked to assume. How will the changing publication and communication landscape impact the library's functions? And how in turn, will libraries respond? Erosions of responsibility in some areas are often compensated by opportunities in other areas. Let us examine some traditional library roles such as creating collections and access to those collections, and how those roles may evolve in the coming years.

One of the most basic functions of a library is collection development, acquiring materials from publishers and other sources to meet the academic research and teaching needs of their campuses. However, libraries are now are moving 'backwards' into a publishing role through experiments in building and supporting institutional repositories of faculty publications. While e-repositories are not ubiquitous, they are well along the way to institutionalization. In fact, MIT's development of D-Space and the University of Southampton's EPrints.org[12] are predicated on the assumption that universities and other organizations need a suite of technical implementation tools and best practices to help them as they collect, organize, and make available the scholarly output of their faculty. Consensus is growing that the university – and often the university library – should play a leadership role in providing 'publishing space' for scholars[13]. Providing this publishing space can be defined as an extension of that traditional collection development role. By managing institutional repositories, libraries are collecting the intellectual products of their faculty. This is electronic collection development at a finer level of granularity than libraries have ever done, intentionally, in the past.

But there is as yet no consensus about who should take the lead in ensuring persistent, efficient, and useful access to these scholarly materials. Universities and their libraries are recognizing that institutional repositories require some form of structured information about the documents that is made publicly accessible in a standardized way.[14] As libraries assume a greater role in publishing with e-repositories, they must define the purpose and extent of their involvement in these repositories. Should they simply provide a warehouse or should they build sophisticated subject access to the publications within their repositories—in essence, performing a value-added aggregator role?[15] Or, conversely, might this subject access role of aggregators—or at least that part that libraries play—become unnecessary? Perhaps software could become sufficiently sophisticated that it is able to perform that aggregation function for a field autonomously. Perhaps software could even provide services similar to what a library does in crafting collections to support an academic department's research needs or what the American Psychological Association does with *Psychological Abstracts* to 'collect' and provide access by an academic subject?[16]

Not only might software replace these sorts of aggregation functions, but the author could replace many of the cataloging and indexing functions at the point of inception. These burdens may be falling more heavily on the information creator through ideas such as author-supplied metadata. Could authors evolve into self-publishers and self-catalogers thus rendering commercial publishers and libraries obsolete? Would authors be willing to play an active or a passive role in metadata creation and indexing? Is it likely that authors would take the time to supply enough information? If there were sufficient data in each publication or information object to harvest, would there be any need for additional human mediation beyond what the author creates initially? How could the inevitable corrections or changes be effected? How would standardization of search elements and terms across distributed repositories be controlled? Building on their past collaboration with faculty, their experience in 'metadata creation (aka, cataloging), and their new e-repository role, libraries could be leaders in helping authors who found themselves in their new roles of document meta-data creators.

Self-publishing and indexing of scholarly e-publications will take, in the most optimistic view, a multiplicity of solutions. Some technical developments, such as XML, seem on the brink of providing a partial solution. However, software aggregation and structured authoring are only two ways of approaching the problem. It is likely that solutions will need to be tailored, especially in this inchoate period, to the authors' readiness, the technology, and the end result. Because libraries have subject specialists who have solid experience working with the scholarly authoring, researching, and educational communities, they can be extremely successful in partnering with content producers and users to create the access to this new breed of publications housed in e-repositories.

Providing information access and helping information seekers are again basic library functions. What role should libraries play in providing subject access to what has been characterized as the 'deep Web' of scholarly electronic literature?[17] Some libraries are beginning to explore this with the institutional repositories that they are creating. An example is the University of California's eScholarship initiative. Its repository "provides persistent access and makes the content easily discoverable".[18] The author agreement commits the Repository to creating a full bibliographic entry for each item deposited, and one of the benefits described on their Website is sophisticated searching.[19] A variety of other experiments with subject indexing are starting, such as those registered with the Open Archives Initiative as 'service providers'.[20] Through institutional and subject-based repositories, there is an increasingly available body of electronically published scholarly materials which can provide the raw material for experiments in subject access.

But, libraries are already involved in simply building these electronic repositories. Could they also assume the additional burden of leadership and cost to develop sophisticated search systems for these repositories? There are rarely simple or completely cost-free solutions to such complex and broad problems. John Ewing, Executive Director of the American Mathematical Society is rightly wary of "miraculous solutions to previously intractable problems…at no cost to anyone".[21] How expensive might search systems be? In an era of severely declining budgets, the need to take on yet another (potentially costly) leadership role is not what academic library managers wish to hear. Neither is it what library specialists in collection development, metadata cataloging and retrieval, or reference services wish to have added to their already overflowing job descriptions. However, libraries, which are masters at both understanding information needs and mediating between researchers and third-party information producers, are in an excellent position to help define the context and outcomes of experiments in this infant area of subject access to the scholarly e-literature.

Libraries also have extensive experience with setting standards, collaborating to create shared cataloging, and listening keenly to the information needs of their user communities. If any group is uniquely positioned to provide subject-specialized organization and access to the scholarly deep Web, academic libraries are. The cost of taking on this function may not be as prohibitive as it could appear at first. Not only are many new experiments trying to develop cost-effective alternatives to labor-intensive cataloging, indexing and abstracting, there are some substantially successful current models already functioning.[22] Studying such ongoing efforts will provide useful data and experience that can be applied to further experimentation.

## SPIRES Collaboration as a Model

In this paper we describe such an effort, which has been working for approximately thirty years. It has helped lead the transition from a totally print-based system to an almost totally electronic-based system. In the process, it has expanded to provide worldwide subject-specialized access not only to the field's journal literature, (as do database vendors), but to a wider set of information objects comprising a significant amount of the intellectual 'ecology' of the field.[23] This is not an effort that is particularly well-funded. In fact, it operates only through a careful use of every (automated, cost-lowering) software program it can implement, a judicious use of hands-on intellectual oversight and cataloging, an aggressive commitment to collaborative and consortial information sharing, and – most radically and uniquely – the volunteer efforts of many of our users.

The SPIRES High-Energy Physics databases provide access to the literature, people, institutions, research, and experiments in the fields of particle and astroparticle physics. First invented and developed by the Stanford Linear Accelerator Center (SLAC) Library in 1974 to acquire, catalog and provide access to high-energy physics pre-prints (advance copies of papers submitted to journals), it is now managed and developed by an international collaboration of laboratories and universities, with substantial volunteer assistance from publishers and researchers. In 1975 an average of 70 papers per week were added to the Research Literature database by the SLAC Library staff. In the first six months of 2003 an average of 700 papers per week were added. The core work of content identification, data entry, subject/access point indexing, authority control, and URL linking, are performed through a blend of software and humanware. We have estimated that, worldwide, there are currently approximately 12 'people' (full time equivalents) dedicated to the work of building these databases. This number contrasts with an estimate of approximately 5 total 'people' who worked on the databases in 1975.

Combined, the six core databases (research literature, experiments, conferences, institutions, people, and jobs) contain about 700,000 records. However, this type of statistic is not a full reflection of the complexity and depth of the information available in and through the databases. In the Research Literature database, for example, one bibliographic record may contain, for example, 150 unique searchable elements, as well as links to a variety of other distributed information such as full text published and unpublished versions, abstracted data, reviews, conference websites, and experimental information. A typical record for a theory paper (which tend to have relatively few authors) is shown in figure 1.

SLAC | Library | Conferences | Experiments | Institutions | Hepnames | Other Databases

Questions/Comments
Help
Search HEP

**SLAC SPIRES HEP**

Searching for papers published in Physical Review Letters (PRLTA)

FIND A MIRABELLI AND J PHYS.REV.LETT.

Browse Author | Format: Cites   Citesummary   LaTeX

Papers 1 to 1 of 1

COLLIDER SIGNATURES OF NEW LARGE SPACE DIMENSIONS.
By Eugene A. Mirabelli, Maxim Perelstein, Michael E. Peskin (SLAC). SLAC-PUB-8002, Nov 1998. 10pp.
Published in Phys.Rev.Lett.82:2236-2239,1999
e-Print Archive: hep-ph/9811337

TOPCITE = 100+
  References | LaTeX(US) | LaTeX(EU) | Harvmac | BibTeX | Keywords | Citation Search
  Abstract and Postscript and PDF from arXiv.org (mirrors: au br cn de es fr il in it jp kr ru tw uk za aps lanl )
  SLAC Document Server
  Phys. Rev. Lett. Server
  CERN Library Record

Browse Author | Format: Cites   Citesummary   LaTeX

Papers 1 to 1 of 1

FIND A MIRABELLI AND J PHYS.REV.LETT.

Search HEP | Other Databases | Conferences | Institutions | Experiments | Help | Comments
SLAC | SLAC Library | Stanford University

Stanford Linear Accelerator Center   Questions and Comments to library@slac.stanford.edu
Location: SPIRES HEP Results Display

In the Experiments database, each record for an experiment contains the equivalent of a multi-page 'encyclopedic' entry which describes the scientific proposal, lists all the experiment members and their institutional affiliations (many experiments have hundreds of scientists), includes some of the past history of that experiment, and provides a comprehensive, up-to-date bibliography of its publications.

Recently a Nobel Laureate in physics, writing about the SLAC Library and the research databases it manages, said, "Over the years its cutting-edge systems and services have helped transform the way we do research in our field."[24] How can a library have such a profound effect? With our secret weapon—our users! There are several broad ways in which users have collaborated with us over time: quality control of the information in the databases; collection development and collection creation; software development, and pure creative genius.

One of the most traditional instances of user quality control is common in many academic libraries. How many of us have had an irate faculty member point out a catalog entry which has that individual's name, affiliation, or work displayed incorrectly? We are typically alerted to such errors by authors emailing us. Because we are a leading information resource in the field, our world-wide users perceive it as important that our information about their publications be correct. Thus a "bootstrap" effect is at work here, as we become important to the field, it becomes easier to maintain good data because we receive more help. We also make use of their interest in having correct personal data by asking them to review their entry in our directory of people in particle physics. We run an automated program periodically that requests that each person in our directory database review the data we have about that person and let us know if it is current and correct. From the replies of authors and researchers we are able to fairly painlessly update this directory of approximately 40,000 entries. This database, while a useful resource itself, also helps build our name authority control system.

In addition to authors pointing out errors in the bibliographic information about their own works, we also fairly regularly receive emails from users who point out typographical errors in the bibliographic entries for works which they did not author. Most frequently, we receive emails pointing out the omission of a particular citation from the list of references for a paper in the research database. While some errors are ones our automated system or human review didn't catch, other errors are made by the original author, for example, while citing someone else's work.[25] After receiving such an email, we check the cited reference against other instances of it in our database and correct, if needed, the author's mistake. Our ability to catch citation errors means that we can correct trails of errors that have developed over time. Errors may accumulate because an author re-uses older reference lists, and so an error once made is inadvertently repeated. They also may develop because another author cuts and pastes from a colleagues' paper and adds to the reference list s/he is developing. Even if authors have read the original papers, they are very unlikely to compare the citations with the reference list, and thus can easily propagate an incorrect reference through many papers.

Our reference lists, then, can be more accurate than those of the original papers. To make use of this we have developed a way for authors to build their reference lists directly out of our research database in a format that can be simply and efficiently added to their paper. Commercial products, such as EndNote and ProCite, permit this kind of downloading and formatting also. This saves the author from the tedious business of reformatting citations to meet a particular journal's editorial requirements and primarily functions as a service to our users. However, in the markup language, we have buried data that makes the processing of that list now far more automatic than reference lists that are not pre-searched. This enables us to save tremendous amounts of staff time reviewing error lists of non-matched references.

A related service we provide to our users is automated reference checking for a list of references an author sends to us via email. An author submits a paper's bibliography to be matched against our database. Then, if the bibliographic information matches, the author knows there are no typographical errors in the new paper's reference list. Non-matching entries are highlighted and the author is alerted to check them for errors. In this way, we help the authors' quality control of their papers and ensure that the reference lists which are eventually added to our database from those papers are correct.

With these volunteer opportunities, building references lists from our database and checking reference lists against the database, users are taking actions that they would do, perhaps with a slight extension or variation, as part of the normal authoring process. Authors have to list and format the works they reference in their papers. They also (we hope!) have to check those lists for typographical errors. By giving them a way to perform both of these functions via systems that help us, there is mutual benefit. We trade their ease of getting or checking references for reference lists that we can process quickly and accurately into our database. When building systems that plan to rely on effort from authors beyond the traditional boundaries of writing a paper, there has to be some direct incentive for them to change or take on additional tasks. The benefit of the community in general is often not a strong motivator, while direct personal benefit in terms of saving time and effort, will alter behavior quickly.

Another area of quality control our users participate in heavily is in catching citations that were added after the e-print was posted and before the paper is published in a journal. One informal study estimates that about thirty percent of e-prints have some substantial revision (not simply typographical changes) but a change in wording, data, or papers added to the reference list before they are published in their final version in a journal.[26] Our cataloging begins with the e-printed paper and so the bibliographic data that we process comes from that version. We have automated systems that compare core data, such as the title and author lists between the unpublished and published versions. However, comparing reference lists for additional citations added between posting to arXiv.org and publication in a journal is not sufficiently automated that we can afford to repeat this procedure on all papers.

Again, we rely heavily on users to help us with this—either the authors themselves or, often, the people whose own works were added to that reference list after the e-print was posted to arXiv.org. Without help, we could replace the draft reference list with the completed one in proportionately few papers. For many papers we have no automated process to replace the e-print reference list with the published version, and so, at present, the additional labor to identify and replace these reference lists would be prohibitive. Some journals send us the reference lists of papers they have published and, again, we are able to replace the draft lists in those circumstances without additional labor. In some cases, the author sends us a new reference list to replace the draft list using a Web form that automatically formats the bibliographic information to fit our database. To help users with this, we have developed a Web page which they can use to send us references that were omitted from a paper.  When using this form, the data can be put into the database with no additional keystrokes from our staff.  Typically, users employ this form to send us an omitted citation when it is their work that was added after the e-print version was posted at arXiv.org.  They have an additional motive for doing this since an updated reference list that includes their work leads to a higher citation count (and greater glory) for the user.  Our thousands of users provide a much-needed additional set of eyes and typing fingers!

Occasionally users will send a full reference list via our Web form for a paper that is in the database but lacking references.  A reference list may be left of, for example, because the paper was neither e-printed nor published in a core journal in the field. For such non-central publications, we do not have the manpower to create reference lists manually. .  In the majority of the cases, there is a measure of self-interest involved in sending a full reference list to us. The paper without a list is usually one either they've authored or that cites a paper they've written.  But we do have other volunteers who send lists for papers which do not include them in any way and which they simply wish to have more fully represented in the database.  The most active example of this activity is a user who sends reference lists from hundreds of papers where he is neither an author nor cited by the author.  In this volunteer's case, there is no self-interest involved, simply a lot of work he performs *gratis* to help improve the database content. In another helpful user's case, he has said he feels obligated to balance requests involving his own works with more altruistic error corrections. In typical physics tradition he quantified his effort at a 10% personal to 90% communal ratio. We need to research user motivation further to see if or how we could encourage such volunteerism more broadly.

There is another area of complex information in the research database that is improved in quality from user cooperation.  Papers written by an experimental group can have a large number of authors.  Anywhere from 50 to 800 authors, along with each author's institutional affiliations, can be listed on a single paper.  For the particle physics community it is important to track all the author names and to link those names with each author's institutional affiliations.  Generally, authors have one or two institutional affiliations, the university at which they work, for example, and the laboratory at which they have experimental privileges.  This can make for a complex and error prone 'author field'.   With some large experimental groups, we have co-developed a system where the experimental group scientific publication coordinator sends an electronic file of each paper's author/affiliation list formatted for automatic input. Since the hundreds of names on these lists may change from one paper to the next, such user-generated input ensures a high level of quality control without our time editing or comparing records. Again, this system works well if the experimental group is well motivated (i.e., sees SPIRES as an important information resource) and conversely tends to fail if this motivation is absent.

Another way that users participate with us to improve the quality of database content we offer is to let us know when we have missed a paper, a conference, an institution or an experiment they believe should be included. Many libraries encourage their local faculty to recommend books or journals for their collections. We receive around thirty of these emailed recommendations per week alerting us to a new experiment that's been formed or to articles that are often on the edges of the fields we traditionally cover. Many of these come via a simple email but we do have a Web page that prompts the user for the bibliographic data elements of an omitted article in a structured way and formats that information so it can be added with little human effort. Some of our users are using this form to prepare, in essence, cataloging records, for papers that we've missed.

Besides correcting errors and omissions, our user community helps us in many other ways. They create content or software that adds value to what we offer. Our user community is incredibly inventive, extremely computer literate and unabashedly assumes they can 'do anything'. Particle physicists have a long tradition of building computer 'hacks' to make their lives easier. SPIRES is, in some sense, a repository and a beneficiary of that habit of hacking. Often they invent something that turns out to be an extremely useful tool, product, or service that integrates well with our databases and mission. At times they have invented something directly for us, but at other times, they have been middlemen for us with other innovators, helping us to take advantage of inventions as they are being developed. The result of this support from our users has been that, at times even during the initial development phase, those inventions have been optimized for our needs.

One example of a user building content was the creation of a directory of people involved in particle physics research. A physicist approached the SLAC library with the idea for this reference tool. The librarian built the database structure and the physicist worked to populate it with content. One of his strategies was to persuade colleagues at other physics departments, schools and institutions to send him regular feeds of electronic records of their scientists and staff. He then wrote programs to help bring that data into the SPIRES database. While the SLAC Library has coordinated the management of this database, called HEP Names (High-Energy Physics), there has been, over time, a series of physicists who have taken the lead role in building its content.

Sometimes our users create a system or develop content that provides an additional level of analysis about the information residing in a database. Two examples of this of this are the software programs that analyze citations or consolidate citation rates for an individual author. We count how often an e-print and its subsequent appearance as a published paper are cited by other authors. A Japanese physicist wrote a program that analyzed citations for all of an author's papers and produced a citation summary displayed graphically. Originally he did this as part of a broader analysis he was doing for the Japanese government on the impact of Japanese science. He and the SLAC library both recognized that it could be a very popular addition to our suite of services and it was installed at SLAC but maintained by him remotely for many years. Recently another user of the databases has sent us software he wrote for us to update the original analysis program.

We also have 'power users' who participate in advising us on and helping us with the databases on many levels. Just one example of this level of user is a particle physics theorist. Using an analysis we produce yearly which ranks the top-cited papers for the past year and the all-time top-cited papers in the entire research literature database, he writes an annual review article that discusses the papers and explores the trends in particle physics based on these citation counts. The 'all-time' highest cited papers average roughly 100 articles that have received more than 1,000 citations recorded since 1974 when we began to track citations in the database. This annual review of 'top cites' is the most popular and eagerly awaited content in the database. He also edits "A SPIRES Guide to the Review Literature in High-Energy Physics" which organizes thousands of review articles into subjects and into further subfields. This bibliography is a particularly valuable teaching tool since it provides quick access not only to the review literature, but full text access to the review itself as well as to all the research publications which the review examines.

A profoundly important example of SPIRES users acting as middlemen between SPIRES and other services grew out of the feedback a theoretical physicist, Paul Ginsparg received when he invented the e-print archives, originally called xxx.lanl.gov, now arXiv.org. When he created this automated repository of electronic versions of preprints, he thought it would be a good way to eliminate the cost and waste of the previous tradition of physicists sending out advance copies of their papers (pre-prints) to their colleagues for discussion. He wanted to eliminate the enormous amount of paper and postage involved, and to create a system that was easy to use to 'post' a paper and would provide broad and convenient access to these advance copies. Since, at the time, a paper submitted for publication to a physics journal took an average of a year to appear in print, having an almost instantaneously available electronic copy was an incredible innovation.

Ginsparg built the archive system and then asked his colleagues to test it out for him. Enthused by the innovation, scientists at SLAC told SLAC librarians about it immediately and encouraged them to develop a connection between the electronic full text papers in the archive and the SPIRES bibliographic database. This led to a collaboration in which the SLAC librarians advised him about what minimum bibliographic information to require from authors (perhaps this was the first electronically author-supplied metadata?). In turn, the SPIRES research database began putting e-print archive identifying numbers into the bibliographic records. Perhaps even more significant for individual authors who posted e-prints, SPIRES began to include citations to the e-print version of a paper in that paper's total citation count.

From this collaboration, the SPIRES system gained the ability to download nightly both the bibliographic data authors supplied and the electronic versions of the reference lists. This enabled a record for a paper to appear in the database literally within hours of the paper first appearing "in e-print". We also shared key elements of the data we added into the literature database with the e-print archive so that their records could display the list of cited references (linking back to our database) and could have information about where an e-print was eventually published. The e-print archive of electronically accessible papers created a repository of data that we could bring into the literature database far more quickly and with less human intervention than previously.[27]

Another innovation that saved SPIRES labor costs was a software program written by a Brown University physicist that converted documents written in the TeX text formatting system, the ubiquitous authoring tool in particle physics, into postscript so that they could be easily printed or displayed on a computer screen. In order to catalog fully the e-print papers downloaded from arXiv.org, the SLAC Library was ftp-ing from the archives TeX papers and converting them to postscript. Scientists who heard about this software recognized that an automated approach could bring the SLAC library significant labor savings and helped persuade the physicist to install his software at SLAC where he continued to maintain remotely for several years. Eventually, the staff at arXiv.org took over the software processing producing postscript for viewing or printing directly from the TeX files on the archive server. Both the original software and the assumption of the TeX to postscripting function by arXiv.org saved the SLAC library a great deal of time spent obtaining eye-legible copies of the e-prints.

One of the most profoundly important examples of an innovation created by a physicist which was then used to improve the services we offer was the World Wide Web. Tim Berners-Lee, the inventor of the Web, had invited a SLAC computational physicist, to see a very early demonstration on his NeXt computer which was at CERN, the European high-energy physics laboratory in Geneva, Switzerland. The SLAC physicist almost did not make the time to go to Berners-Lee's office, but did and recognized immediately that this program could become a useful interface to the SPIRES particle physics databases.[28] He brought a copy of the program back with him on a floppy and enthusiastically showed it to the SLAC librarian, insisting that it could be the Internet search interface for which the SPIRES databases had been waiting. With the help of other physicists and programmers who volunteered their time, they had the first website in the U.S. up and running within a couple of months and were writing html out of the database on-the-fly. When Berners-Lee learned that his software was the new search interface to the SPIRES research database, he was delighted, saying that this would be the 'killer app' for his invention.[29] And, in fact, it was by using SPIRES through the Web interface that the first Web user community, particle physicists, was born.

The history of the SPIRES databases and their continued successful existence is intimately connected with the contributions—both large and small—that users make to the databases' quality, content, and continued innovation. But, is this simply the story of an isolated set of behaviors by a unique and unreplicatable community? What lessons, if any, can be learned that might be of use to the future development of the scholarly publishing and communication process?

## Conclusions

**Lesson One:** Users need and value in-depth subject access. This fact is shown by the willingness of users to help us at all, as well as by our use statistics.[30] Some emailed comments from users around the world express how central they feel the literature database is to their research.

*"Thank you so much for what you are doing. You are building the greatest science library in the history of mankind..."*

*"Thank you for the eminently useful service you provide for the HEP community."*

*"The virtual library [the SLAC Library] provides to the worldwide particle physics community levels the scientific playing field"*

Their willingness to spend time helping in so many ways is based on their need for and appreciation of the databases. In particular the kinds of software development, content development and mediating between other innovators and SPIRES are motivated by this broader loyalty to the SPIRES system.

**Lesson Two:** With volunteer efforts, particularly continued automation improvements, and a distributed approach to building the resources, in-depth subject specialized databases are not prohibitively expensive. Because SPIRES is seen as important to the field, users are motivated to help us. What might start as a relatively small and expensive project, particularly until use grows, can turn into a very helpful, appreciated service, by this 'bootstrap' effect. Of course, the timescale needed to build 'brand loyalty' is not always short plus there may be significant front-end costs during the start-up period. However, one motivation that does certainly contribute to success is the feeling of ownership among the user community. By encouraging and using volunteer efforts we have made the users feel as though SPIRES is not a third-party service, but instead an extension of particle physics, as indeed it truly is. This encourages the type of loyalty and helpful spirit that permeates our user community today.

**Lesson Three:** There are successful ways to motivate authors to take an expanded role in the scholarly publication and communication process beyond their core efforts of research and writing. It is clear from our experiences that enlightened self-interest is a strong motivator for author participation. Giving authors some carrot or reward works effectively. Additionally, getting them to make minor changes in a part of their writing or publication process also seems to work. While some users are motivated by the long-term benefit to the field in general, this is the exception not the rule. Most authors do not want to take the time to format their documents, or add metadata to a database, unless there is some clear benefit to them. For this reason third-party services may always be useful, and services that do rely solely on author-supplied information, need to think carefully about the motivations of the authors doing the work. There needs to be a 'fair trade' of benefits between the author and the information service.

While many parts of the scholarly publishing and communication structure may emerge, change, or grow moribund in the next decade, researchers and students will continue to need persistent and consistent access to scholarly literature. While its exact future may not be clear, there is a clear continuing need for collection selection and access, in essence, that aggregator function. While libraries may not become the exclusive providers of this service they should take a leadership role in ensuring that the best possible systems for scholarly access are developed through partnerships with other players in the publication and communication system. The SPIRES consortium's system of identifying relevant research, data, and other information objects, and enabling in depth subject access to that body of information via a sophisticated suite of databases and services provides a model of in-depth support of scholarship at a cost-effective level. The SPIRES system is a prime example of the utility that libraries can continue to provide in an increasingly electronic environment. Perhaps more importantly, the SPIRES experience shows that collaboration with authors, users, and others in the academic community is not only possible but essential if one is to build collection and access systems that continue to evolve to meet researchers' information needs an increasingly e-Research world.

## Acknowledgements

We are grateful to Ann Redfield and Louise Addis, and our colleagues in the SLAC Library for discussions and assistance with the issues presented here.

---

[1] Brian L. Hawkins, "Information Access in the Digital Era: Challenges and a Call for Collaboration" *EDUCAUSE Review* 51 (Sept/Oct. 2001), p. 54.

[2] Charles W. Bailey, Jr., *Scholarly Electronic Publishing Bibliography*, Houston: University of Houston Libraries, 1996-2003. In this bibliography, "Chapter 7: New Publishing Models" lists 143 articles calling for, describing or analyzing the evolving publishing revolution. Available from http://info.lib.uh.edu/sepb/models.htm.

[3] Paul Jones, "Open(Source)ing the Doors for Contributor-Run Digital Libraries" *Communications of the ACM* 44, no.5 (May 2001): 46.

[4] Several authors have addressed the problem of how much self-cataloging and publishing management to expect of scholarly authors. Erik Dugan, et al., "The Ariadne Knowledge Pool System" *Communications of the ACM* 44, no.5 (May 2001): 73-78. This article is a useful introduction to the issue of when author-supplied metadata is best captured and the need for additional third-party intervention. Jeffrey R. Young "'Superarchives' Could Hold All Scholarly Output" *The Chronicle of Higher Education* 43 (July 5, 2002): A29.

[5] Representatives Sabo, Kaptur, and Frost introduced a bill into the House of Representatives on June 26, 2003, H.R. 2613, the 'Public Access to Science Act', which proposes to "amend title 17, United States Code, to exclude from copyright protection works resulting from scientific research substantially funded by the Federal Government". Available from http://frwebgate.access.gpo.gov/cgi-bin/getdoc.cgi?dbname=108_cong_bills&docid=f:h2613ih.txt.pdf

[6] Peter J. Denning and Bernard Rous, *The ACM Electronic Publishing Plan*, (1995), p. 4. Corrected version available from the *Communications of the ACM* http://acm.org/pubs/epub_plan.html.

[7] Unpublished presentations: "Peer Review in the Age of Open Archives", May 24-25, 2003. Sponsored by the Interdisciplinary Laboratory of the International School of Advanced Studies (SISSA), Trieste, Italy. Contact workshop organizers: Marco Fabbrichesi (marco@he.sissa.it), Stevan Harnad (harnad@ecs.soton.ac.uk), Stefano Mizzaro (mizzaro@dimi.uniud.it), and Corrado Pettenati (corrado.pettenati@cern.ch). Marco Fabbrichesi and B. Montolli, "Peer Review: A Case Study". Available from http://tips.sissa.it/docs/peer_review.pdf.

[8] Discussions held Jan. 16-18, 2001 at SLAC with Stewart C. Loken, et al. APS Task Force on Electronic Information Systems. (Report in draft form, contact: scloken@lbl.gov.

[9] Available at http://www.iop.org/EJ/S/UNREG/EGeUUag1CjHJf25bGOmSIA/ejs_extra/-coll=becm

[10] Available at http://www.socscinet.com/evaluation/index.html